\documentclass[runningheads]{article}
\usepackage{amsmath}
\usepackage{amssymb}
\usepackage{booktabs} 
\usepackage{caption} 
\usepackage{subcaption} 
\usepackage{graphicx}
\usepackage{mathrsfs}
\usepackage{appendix}
\usepackage{bigints}
\usepackage[version=4]{mhchem}
\usepackage{mathtools}
\usepackage{pgfplots}
\usepackage[all]{nowidow}
\usepackage{tikz}
\usetikzlibrary{er,positioning,bayesnet}
\usepackage{multicol}
\usepackage{algpseudocode,algorithm,algorithmicx}
\usepackage{natbib}
\usepackage{hyperref}
\usepackage[inline]{enumitem} 
\usepackage{authblk}
\graphicspath{ {figures/} }

\definecolor{blue}{HTML}{1F77B4}
\definecolor{orange}{HTML}{FF7F0E}
\definecolor{green}{HTML}{2CA02C}

\pgfplotsset{compat=1.14}

\begin{document}
\title{Impacts of peak-flow events on hyporheic denitrification potential}

\author[1]{Tanu Singh \thanks{Corresponding author: tanu.singh@tum.de}}
\author[2]{Shubhangi Gupta}
\author[3,4]{Gabriele Chiogna}
\author[5]{Stefan Krause}
\author[1,6]{Barbara Wohlmuth}
\affil[1]{Department of Numerical Mathematics, Technical University of Munich, Garching, Germany}
\affil[2]{Marine Geodynamics, GEOMAR Helmholtz Center for Ocean Research Kiel, Germany}
\affil[3]{Department of Hydrology and River Basin Management, Technical University of Munich, Germany}
\affil[4]{Institute of Geography, University of Innsbruck, Innsbruck, Austria}
\affil[5]{School of Geography, Earth and Environmental Sciences, University of Birmingham, Birmingham, UK}
\affil[6]{Department of Mathematics, University of Bergen, Bergen, Norway}
\date{}
\maketitle             
\begin{abstract}
Subsurface flows, particularly hyporheic exchange fluxes, driven by streambed topography, permeability, channel gradient and dynamic flow conditions provide prominent ecological services such as nitrate removal from streams and aquifers. Stream flow dynamics cause strongly nonlinear and often episodic contributions of nutrient concentrations in river-aquifer systems.  Using a fully coupled transient flow and reactive transport model, we investigated the denitrification potential of hyporheic zones during peak-flow events. The effects of streambed permeability, channel gradient and bedform amplitude on the spatio-temporal distribution of nitrate and dissolved organic carbon in streambeds and the associated denitrification potential were explored. Distinct peak-flow events with different intensity, duration and hydrograph shape were selected to represent a wide range of peak-flow  scenarios. Our results indicated that the specific hydrodynamic characteristics of individual flow events largely determine the average positive or negative nitrate removal capacity of hyporheic zones, however the magnitude of this capacity is controlled by geomorphological settings (i.e. channel slope, streambed permeability and bedform amplitude). Specifically, events with longer duration and higher intensity were shown to promote higher nitrate removal efficiency with higher magnitude of removal efficiency in the scenarios with higher slope and permeability values. These results are essential for better assessment of the subsurface nitrate removal capacity under the influence of flow dynamics and particularly peak-flow events in order to provide tailored solutions for effective restoration of interconnected river-aquifer systems.

\end{abstract}
\section{Introduction}

Multiple physical, chemical and biogeochemical processes occurring in hyporheic zones (the transition zones between  surface water and groundwater) determine the ecological functioning and health of interconnected aquifer-river interfaces \citep{krause2011inter, boano2014hyporheic, krause2017ecohydrological}. The hyporheic zone acts as natural biogeochemical reactor and is crucial for the self-purification capacity of the river-aquifer systems. Several drivers and controls such as channel gradient, streambed permeability, pressure gradient at the sediment-water interface affect the hyporheic flow paths and nutrient transformation and distribution in the hyporheic zone \citep{zarnetske2011dynamics, krause2011hydrology,gomez2014, krause2014understanding}. More recently, we have started to better understand the role of hydrodynamic variability and in particular of peak-flow conditions on hyporheic exchange flow and residence time distributions \citep{malzone2015temporal,wu2018impact,singh2019dynamic,singh2020effects}. Considering the impact of change to these dynamic factors on hyporheic exchange flow and biogeochemical functioning is becoming even more relevant given the increasing frequency, duration and severity of extreme (low and high) flow conditions due to climate change (intense precipitation or snow-melting) or anthropogenic disturbances (e.g., dam operations) and their implications for solute transport, mixing and reactions in the hyporheic zone \citep{bruno2009impact, bruno2013multiple, jones2014dual, casas2015effects, krause2017ecohydrological}.

Catchment nutrient dynamics and their resulting loads in rivers are affected by the hydro-meteorological conditions and runoff generation processes, varying seasonally and also by land use practices, farming strategies, increased urbanization and water resource management \citep{blaen2017high,abbott2019water,lin2021evaluating}. Precipitation events can flush large proportion of nutrients from nearby agricultural lands and activate more distant sources of nutrient "hotspots" that are disconnected and do not contribute to nutrient loads in rivers under baseflow conditions \citep{krause2014catchment}. Episodic peak-flow events can be dominating inﬂuences on catchment nutrient export patterns and dynamics \citep{malcolm2004hydrological,boano_bedform-induced_2007,blaen2017high,khamis2017continuous}. For instance, \citet{lin2019seasonality} showed that the nitrate flux decreased to less than half of spring loads in the higher order stream of an agricultural catchment and \citet{comer2020seasonal} found in the experimental studies that finer-grained streambeds can play a major role in nitrate removal from agricultural catchments. 

Amongst the many biogeochemical processes occurring in the hyporheic zone, detailed understanding of the hyporheic denitrification efficiency is essential to understand the capacity and limitations of the natural attenuation potential of hyporheic zones as solution to high nitrate loads in surface waters and groundwater \citep{findlay1995importance,gomez-velez_flow_2017}. Eutrophication due to excess nutrients (such as nitrogen) availability leads to low dissolved oxygen  concentrations, high turbidity, and low pH with detrimental effects on stream ecology and aquatic ecosystems \citep{smith2009eutrophication}. Besides its impacts on the ecological functioning of interconnected river-aquifer systems, increased nutrient loads do also cause severe effects to socio-economic activities including drinking water supply and recreational activities \citep{bennett2009understanding,abbott2019water}.  Lower order rural catchments dominated by agricultural land use are often important sources of downstream river nutrient pollution, particularly nitrogen. Flow variability can cause increase as well as decrease in nutrient concentrations  in hyporheic zones under different hydrodynamic and hydrogeological conditions.  The interactions between nutrient exchange fluxes, mixing, including travel and reactions times are controlled by geomorphological settings involving channel slopes, streambed permeability and topography. Implications of these interacting dynamics need to be understood in order to make the prediction of the natural capacity of the river-aquifer systems to act as a  natural biogeochemical reactor. Predictive dynamic modeling has been a helpful tool to gain insights into the impacts of peak-flow events not only concerning the improved assessment of resilience in stream ecosystems but also widen the basis to design locally optimized management and restoration strategies. 

Recent advances in understanding the impacts of transient flow conditions on nitrate removal  in river-aquifer systems include \citet{trauth2017single} who investigated the impact of single discharge events on flow exchange, and reactions in the hyporheic zone below a natural in‐stream gravel bar using linear relationship between discharge and nutrient loads. The authors showed that hyporheic exchange ﬂux, solute transport, and consumption increase during individual flow events. \citet{boso2018probabilistic} adopted a probabilistic approach to quantify the impact of uncertainty in both the temporally-variable river water and the rate constants on predictions of $N_{2}O$ emissions at the bedform scale. On the other hand, \citet{zheng2018diel} focused on the effects of temperature on nitrogen cycling and nitrate removal‐production efficiency in bedform-induced hyporheic zones under steady flow conditions. \citet{harvey2013hyporheic} demonstrated that different combinations of local hydraulic and biogeochemical characteristics in these geomorphic units can contribute substantially to whole‐stream denitrification. \citet{gomez2015nature} presented numerical simulations of the Mississippi river network using a hydrogeomorphic model and found that the vertical exchange with sediments, driven by submerged bedforms, has denitrification potential exceeding that of lateral hyporheic exchange with sediments alongside river channels, driven by river bars and meandering banks. \citet{hester2019effect} showed that river stage and sediment heterogeneity pose the strongest control on mixing-dependent denitrification. Despite the above advances, very little is known about the hyporheic denitrification potential during dynamic flow conditions, taking into consideration non-linear observations of nutrient concentrations and under the influence of different geomorphological settings.

To address this knowledge gap, we develop robust flow, solute, and reactive transport models for simulation of transient two-dimensional reactive transport to estimate the hydrodynamic impacts of hyporheic denitrification potential during peak-flow events. We evaluate the impacts of peak-flow events by taking into consideration (i) peak-flow events of different magnitude, duration and event hydrograph shape (ii) streambed sediment permeability (iii) stream channel slope (iv) bedform amplitude. Furthermore, we use the analysis of observed stream flow nitrate and dissolved organic carbon concentration relationships across a wide range of conditions to define the scenarios for these simulations. This approach is designed to provide enhanced mechanistic process understanding of impacts of peak-flow events on the fate of nitrate in hyporheic environments, information that will assist both industry and regulators in the design of measures to prevent the disruption of self-purification in river ecosystems.

\section{Materials and Methods}

A physics-based model was developed and implemented for performing scenario simulations in this study. The following approach was followed (1) setup of simulation framework including streambed sediment geometry and parameterization of material and hydrodynamic properties; (2) generation of scenario hydrographs using prescribed head distribution across the sediment-water interface; (3) estimation of Darcy velocity; (4) reactive transport modeling of aerobic respiration and denitrification along hyporheic exchange flowpaths using Monod-kinetics; and (5) simulation of model scenarios.
 
\subsection{Description of the model domain}

\begin{figure*}[h]
\centering
\includegraphics[width = \textwidth]{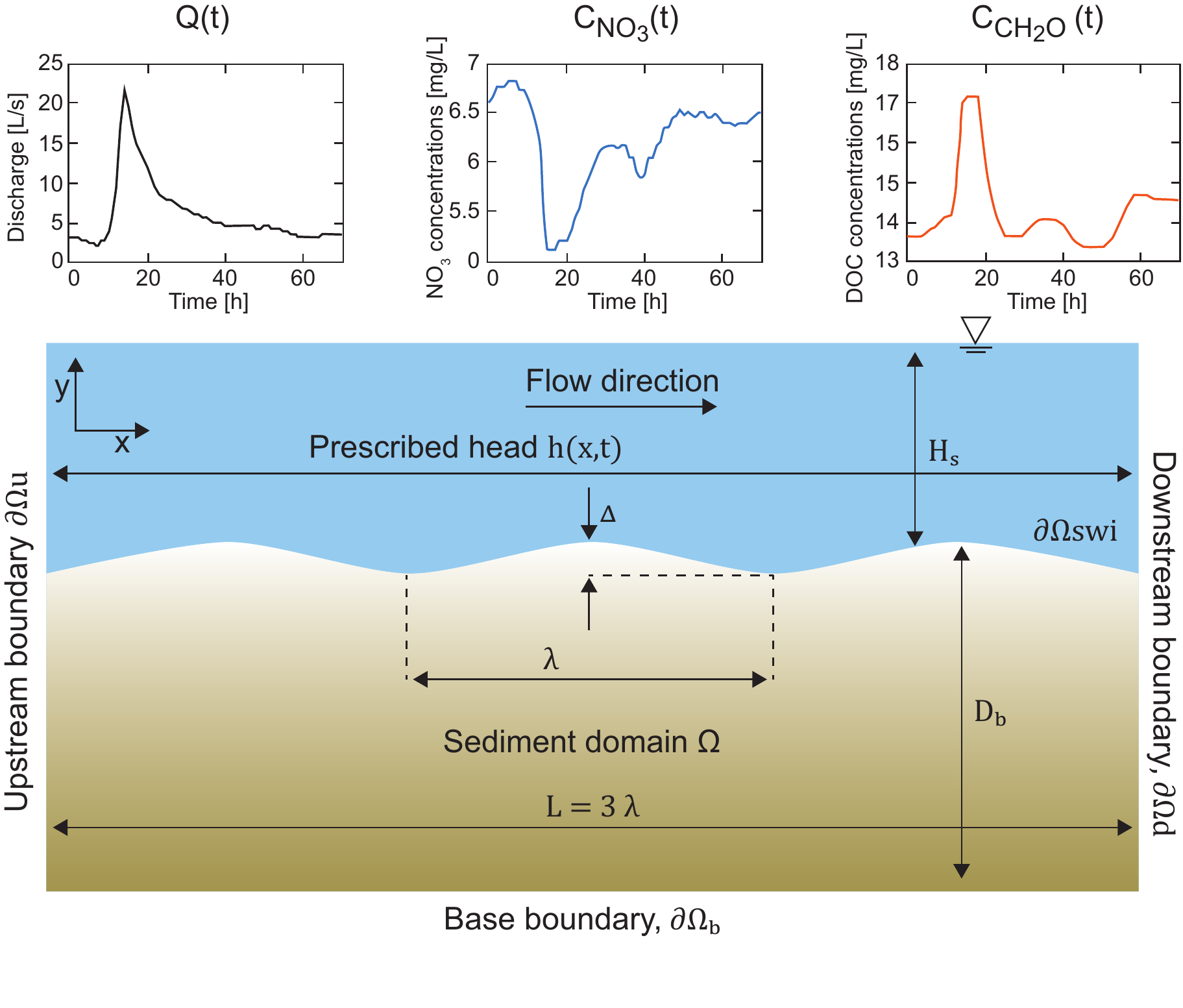}
\par
\caption{Schematic representation of the model. The sediment domain ($\Omega$) is assumed homogeneous and isotropic. Hyporheic exchange is driven by streambed topography, channel gradient and time-varying discharge ($Q(t)$). Prescribed head distribution is imposed along the SWI ($\partial \Omega_{SWI}$). Periodic (pressure drop) boundary conditions are assumed for the lateral boundaries ($\partial \Omega_{u}$ and $\partial \Omega_{d}$) representing an infinite domain, horizontal ambient flow is assumed proportional to the channel slope, and the base of the model domain ($\partial \Omega_{b}$) is assumed impervious. Flow direction is from left to right. Concentrations of nitrate and dissolved organic carbon are temporally variable depending on the event scenarios. }
\label{MethFig1}
\end{figure*}

We use a simplified conceptualization of a streambed-river interface to systematically explore the impact of peak-flow events on bedform-driven hyporheic exchange and reactive transport. For this purpose, we implement transient flow, transport and reactive transport models for porous matrix flow in streambed sediments. The modeling domain ($\Omega$); represents homogeneous and isotropic streambed sediments (See Figure \ref{MethFig1}). It is bounded at the top by the sediment-water interface (SWI; $\partial \Omega_{SWI}$), which is assumed sinusoidal $Z_\mathrm{SWI} = \frac{\Delta}{2}\sin( \frac{2\pi x}{L}) $ where $\Delta$ [L] and $L$ [L] are the characteristic amplitude and wavelength of the bedform, respectively (see Figure \ref{MethFig1}). The total length and depth of the modelling domain are $L_t=3\,L$ and $d_b$ [L], respectively. At the bottom, the model domain is bounded by a horizontal boundary ($\partial \Omega_{b}$) and at sides by the lateral boundaries ($\partial \Omega_{u}$ and $\partial \Omega_{d}$). These dimensions are selected to avoid boundary effects in the numerical simulations.

\subsection{Head distribution at the sediment-water interface and hyporheic flow field}

The pressure distribution at the sediment water-interface ($\partial \Omega_{SWI}$) assumes a linear combination of head fluctuations induced by large- and small-scale bed topography \cite{worman2006exact,stonedahl2010multiscale}:

\begin{equation}
h(x,t) =  - S \, x + H_{s}(t) + \dfrac{2\,h_d(t)}{\Delta} \, Z_{SWI}\bigg(x+\dfrac{L}{4}\bigg)
\end{equation}
where $S$ is channel slope, $H_{s}(t)$ [L] is the time-varying river stage,  $Z_{SWI}(x)$ is the function describing the bed topography [L], and $h_d(t)$ is the intensity of the dynamic head fluctuations \cite{elliott1997transfer},

\begin{equation}
h_d(t)  = 0.28 \dfrac{U_s(t)^2}{2g} \begin{cases} \Bigg(\dfrac{\Delta}{0.34 \, H_{s}(t)}\Bigg)^{3/8}&\mbox{for}\quad \dfrac{\Delta}{H_{s}(t)} \leqslant 0.34, \\[12pt]
\Bigg(\dfrac{\Delta}{0.34 \, H_{s}(t)}\Bigg)^{3/2}&\mbox{for}\quad \dfrac{\Delta}{H_{s}(t)} > 0.34,  \end{cases} 
\end{equation}
where the mean velocity [L$T^{-1}$] is estimated with the Chezy equation for a rectangular channel as $U_s(t) = M^{-1}H_s(t)^{2/3}S^{1/2}$ with $M$ is the Manning coefficient [L$^{-1/3}$T] \cite{dingman2009fluvial}.

Finally, the horizontal and the vertical components of the Darcy velocity can be expressed as \citet{marzadri2016mixing}

\begin{subequations}
\begin{align} 
&   v_x(x,y,t) = v_{0}(t) sin (\lambda x)[tanh(\lambda d_b) sinh(\lambda y) + cosh(\lambda y)] + v_s, \\
&    v_y(x,y,t) = -v_{0}(t) cos (\lambda x)[tanh(\lambda d_b) cosh(\lambda y)+ sinh(\lambda y)] + v_{gw}
    \end{align} 
\end{subequations}
where $v_0 = K \lambda h_d(t)$ is the maximum seepage velocity component due to the bedform morphology [L$T^{-1}$], and $v_s = K S$ is the underflow seepage velocity due to the channel gradient [L$T^{-1}$] and $\lambda = \frac{2\pi}{L}$ is the dune wave number. As we consider neutral groundwater conditions, i.e. the absence of groundwater gaining or losing conditions, we get in our case $v_{gw} = 0$.

\subsection{Reactive transport model}

The reactive transport model for transient reactions \cite{trauth2014hyporheic, trauth2017single} is given by 
\begin{equation}\label{Eq_HZ} 
\theta \frac{\partial c_s}{\partial t} =\nabla\cdot (\mathbf{D}\nabla c_s)   -   \nabla\cdot (\mathbf{q} c_s) + R_s
\end{equation}
where $R_s$ is the reaction rate  (source/sink term) is the solute species s,  $c_s$ is the concentration [ML$^{-3}$] of that species, $\mathbf{q}$ is the Darcy flux [LT$^{-1}$], and $\mathbf{D} = \{D_{ij} \}$ is the hydrodynamic dispersion tensor. 

Infiltration of river-borne nitrate, dissolved organic carbon and oxygen occurs across the sediment-water interface and the following reactions take place
\begin{center}
    Aerobic respiration: \ce{CH2O + O2 ->  CO2 + H2O} \\
    Denitrification: \ce{5CH2O + 4NO3- + 4H+ -> 5CO2 + 2N2 + 7H2O}\\
\end{center}

The reactions are simulated using Monod-kinetics formulation with an additional inhibition term for the denitrification reaction for the conditions when $O_2$ concentrations are high. The highest nitrate consumption occur under low oxygen concentrations i.e. anoxic conditions.  The general Monod-kinetic is in the form \citet{trauth2014hyporheic}

\begin{equation}\label{Eq_monod}
R  = \mu_{max} I \bigg(\dfrac{C_D }{K_D + C_D}\bigg) \bigg(\dfrac{C_A }{K_A + C_A}\bigg)
\end{equation} 
where R is the reaction rate, $\mu_{max}$ represents the maximum reaction rate, $C_D$ and $C_A$ are the concentrations of the electron donor ($CH_{2}0$) and acceptors ($O_2$ and $NO_3$), and $K_D$ and $K_A$ are the half saturation constants of electron donors and acceptors, respectively. An inhibition factor I is required for the denitriﬁcation kinetic, as described by 

\begin{equation}\label{Eq_inhi}
I  = \bigg(\dfrac{K_I }{K_I + C_{O_2}}\bigg) 
\end{equation}

where $K_I$ is the inhibition constant and $C_{O_2}$ is the $O_2$ concentration, while I = 1 for aerobic respiration.  The reaction rate of denitrification is inhibited in the presence of $O_2$, because it is the primary electron acceptor for organic matter oxidation in microbial-remediated reactions. See Table 1 for the full description of how the parameters used for reactive transport have been deﬁned.

\begin{table*}[h!]
\centering
\caption{Input Parameters for reactive transport model \cite{trauth2017single}}
\label{Table1}
\begin{tabular}{p{3.5cm}|p{3.5cm}}
Parameters & Value \\
$K_{O_2}$ & 6.25e-03 mmol/L  \\
$K_{NO_3}$ & 3.23e-02 mmol/L\\
$K_{DOC}$ & 1.07e-01 mmol/L\\
$K_{I}$ & 3.13e-02 $O_2$ mmol/L\\
$\mu_{max,AR}$ & 4.78e-01 mmol/L/d\\
$\mu_{max,DN}$ & 8.64e-02 mmol/L/d\\
\end{tabular}
\end{table*}

\begin{table*}[h]
\centering
\caption[Parameterisation of the numerical model.]{Parameterisation of the numerical model.}
\label{Table3}
\begin{tabular}{p{3.5cm}|p{3.5cm}|p{6.cm}}
Parameters & Value & Description \\
\multicolumn{3}{c}{Constant model parameters} \\
$\lambda$         & 1\,m             & Bedform wavelength\\
$d_{b}$         & 5\,m             & Depth of the domain\\
M               & 0.05            & Manning's coefficient\\
$\alpha_{L}$    & 0.05\,m          & Longitudinal dispersivity\\
$\alpha_{T}$    & 0.005\,m         & Transverse dispersivity\\
$\rho$          & 1000\,kg\,m\textsuperscript{-3}  & Fluid density\\
$\mu$           & 1.002$\times$10\textsuperscript{-3}\,Pa\,s & Fluid dynamic viscosity\\
$g$             & 9.81\,ms\textsuperscript{-2}    & Acceleration due to gravity\\
\multicolumn{3}{c}{Varied model parameters}\\

$\Delta/\lambda$    & 0.1, 0.05    & Bedform aspect ratio   \\
$S$                 & 0.05, 0.01, 0.001    & Channel slope\\
$\kappa$        & 10\textsuperscript{-11}\,m\textsuperscript{2}, 10\textsuperscript{-12}\,m\textsuperscript{2}, 10\textsuperscript{-13}\,m\textsuperscript{2}  & Permeability \\

\end{tabular}
\end{table*}

\subsection{Numerical methods and solution scheme}
\label{subsec:numerical_scheme}

The mathematical model is strongly coupled and highly nonlinear. The model is discretized in space using a locally mass-conservative cell-centered finite volume scheme, and in time using an implicit Euler scheme. The fluxes are approximated using an orthogonal two-point stencil. The discrete model is linearized using a standard Newton method and the resulting linear system is solved using a SuperLU linear solver \cite{superLU99_SEQ}. The numerical scheme is implemented within the DUNE-PDElab framework \cite{Bastian2010} which is based on C++. The numerical simulator is versatile in terms of the geometry of the computational domain, dimensionality of the problem (2D, 3D), and the implementation of the periodic boundary conditions.

\subsection{Scenarios and parametrisation of the model}
\begin{figure*}[h!]
\centering
\includegraphics[width = \textwidth]{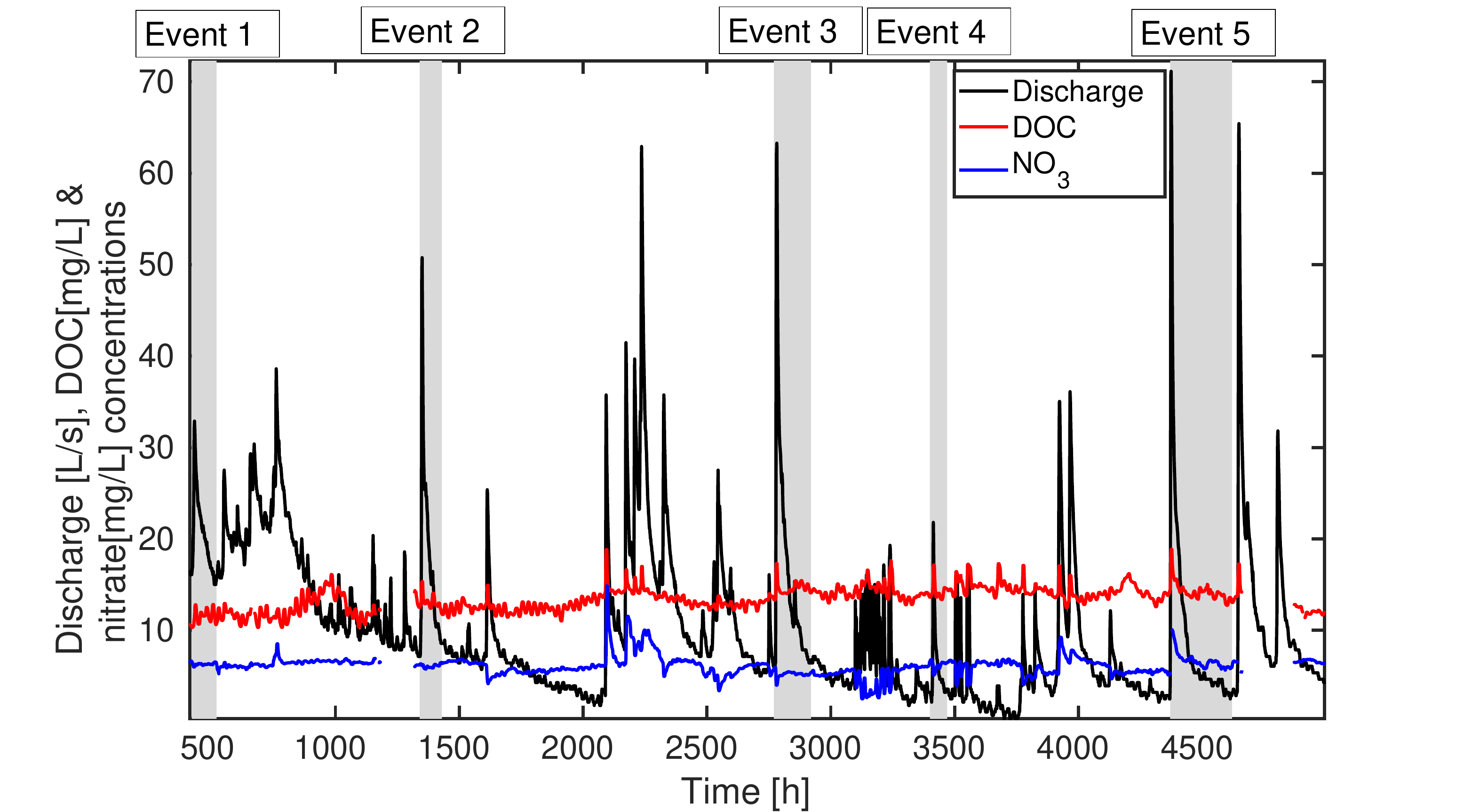}
\par
\caption{Time series of stream discharge, nitrate concentrations, and dissolved organic carbon concentrations. The grey shaded regions denote the selected peak-flow events for the presented study. We use the data presented in \cite{blaen2017high} for the numerical simulations.}
\label{MethFig2}
\end{figure*}
As the relationships between stream flow and solute concentrations can vary significantly spatially between rivers, catchments, geographic and landuse contexts as well as based on the pre-event conditions, we here use the continuous high-frequency observations of discharge and concentrations of dissolved organic carbon and nitrate in an agricultural stream from previous work by \citet{blaen2017high} as an example for demonstrating some of the event dynamics that can be observed in intensively managed mixed-landuse catchments. With the emergence of reliable in-situ water quality sensors high-frequency observations of discharge and nutrients in rivers are becoming more frequent now and extending the scenarios to other site conditions is likely to extend the insights into event-driven nutrient dynamics in hyporheic zones in different systems in the future. Therefore, in this study, high resolution data of streamflow, nitrate and dissolved organic carbon  for the Wood Brook at the Birmingham Institute of Forest Research were analyzed (see Figure \ref{MethFig2}). 

In this study, we considered five example peak-flow events selected for their varying intensity, duration and shape of the hydrograph (see Figure 2 grey shaded regions). The events are selected so that they have isolated shape and consistently  data availability for flow and nutrient concentrations. As studied by \citet{blaen2017high} and see in Figure 2, nitrate concentrations exhibited no clear long-term trend over the entire monitoring period whereas the highest dissolved organic carbon concentrations were observed in late summer. We then performed parametric studies analyzing the effects of varying channel slope, sediment permeability and bedform amplitude of the streambed on hyporheic exchange flow and nutrient turnover for the selected five peak-flow and concentration events .

\section{Results and Discussion}

\subsection{Hyporheic flow field}

\begin{figure}[h]
\noindent\includegraphics[trim=0cm 0cm 0cm 0, clip, width=\textwidth]{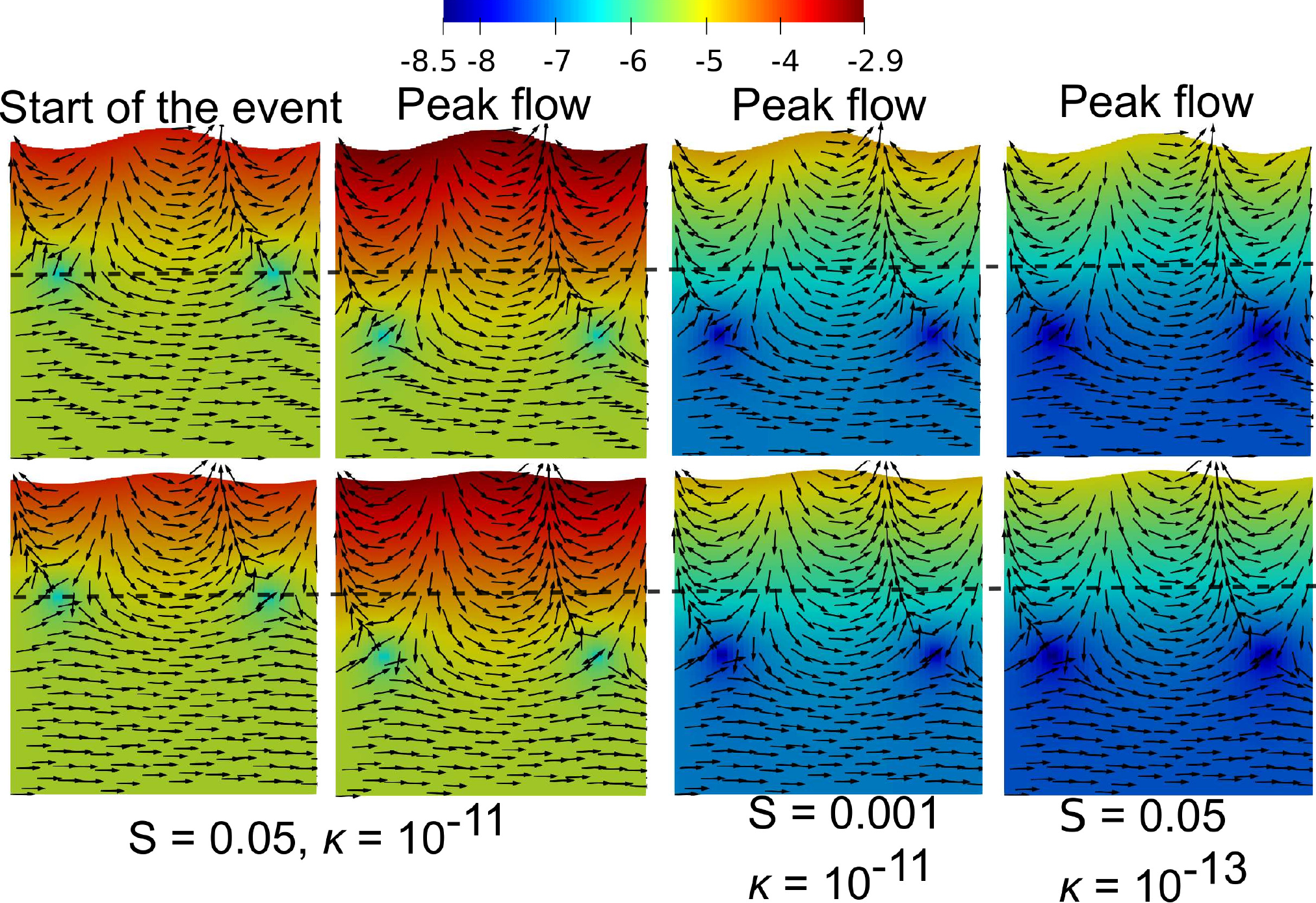}
\caption{Snapshots of the dynamic flow field (black arrows represent direction and are not proportional to the magnitude) in the sediment domain. Surface represents the magnitude of Darcy flux vector (in log scale $m/s$). Top row represents the streambed with $AR = 0.1$ and bottom row with $AR = 0.05$. The first columns are for the times t = 0 (baseflow conditions), second is for peak-flow conditions and for the Event 2 scenarios with $S = 0.05$ and $\kappa = 10^{-11}$, third column is for peak flow conditions and for the scenarios with $S = 0.001$ and $\kappa = 10^{-11}$ and finally fourth is for $S = 0.05$ and $\kappa = 10^{-13}$. }
\label{}
\end{figure}

Figure 3 illustrates the Darcy velocity magnitude in log scale in the sediment domain with variable channel gradient, permeability, and aspect ratio of the streambed. Interplay of factors such as channel gradient (proportional to slope), sediment permeability, streambed topography, and event-induced pressure at the sediment-water interface results in bi-directional circulation systems in the streambed domain (depicted in Figure 3).  This bidirectional flow exchange leads to formation of areas or zones where the velocities are zero or nearly zero. The lower aspect ratio scenario leads to formation of smaller hyporheic zones due to smaller pressure gradient at the sediment-water interface. Similar effect can be observed by the larger channel slope magnitude due to increase in the horizontal ambient groundwater flow from upstream to downstream direction. Larger slopes induce stronger underflow velocities that lead to counteracting force against higher local pressure gradient at the sediment-water interface and prevent hyporheic zones from expanding.  

Stagnation zones are of particular importance in biogeochemical transformations and are often referred to as \emph{biogeochemical hotspots}. The fluctuation of these stagnation points is largely determined by the pressure variations at the sediment water interface. However, the location and the size of stagnation zones are modulated by the hydrogeological properties of the streambed and river flow dynamics. As indicated by the log scale representation of Darcy velocity magnitude, the higher aspect ratios of the streambed lead to stronger velocities and larger hyporheic zones. Similar trends can also be observed for higher permeabilities and higher slope values. As the flow model assumes negligible storage, the pressure variations and propagation are instantaneous within the domain.

\subsection{Spatio-temporal distribution of hyporheic denitrification and DOC consumption}

\begin{figure}
\centering
\includegraphics[trim=0cm 0cm 0cm 0, clip, width=\textwidth]{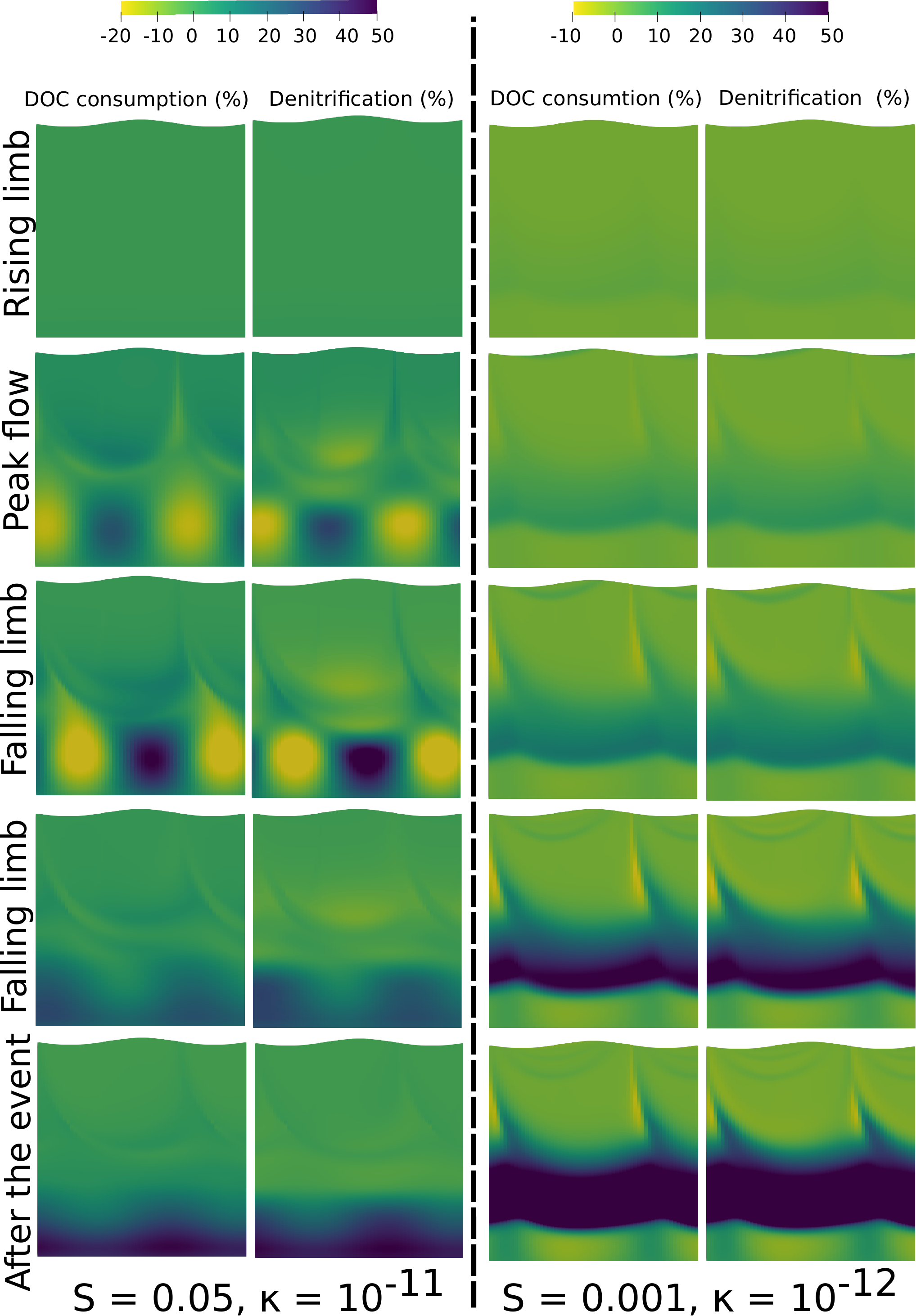}
\caption{Snapshots of the spatio-temporal distribution of denitrification and DOC consumption (\%) compared to the baseflow conditions in the sediment domain for different time points of the Event 2 scenarios. The vertical black dashed line delineates the scenarios with $S = 0.05$ and $\kappa = 10^{-11}$ (left) with $S = 0.001$ and $\kappa = 10^{-12}$ (right). The first row illustrates the reaction rates for early rising limb of the event, the second row for the peak-flow, the third and fourth rows during the recession of the event and finally the  fifth row displays the rates right after the event.}
\end{figure}

Figure 4 illustrates the spatial and temporal distribution of dissolved organic carbon (DOC) consumption and denitrification rates with respect to baseflow conditions for the Event 2 and for the $AR = 0.05$. The black dashed line delineates the scenarios with $S = 0.05$ and $\kappa = 10^{-11}$ with $S = 0.001$ and $\kappa = 10^{-12}$. The first row illustrates the reaction rates for the early rising limb of the event, the second row for the peak-flow, the third and fourth during the recession of the event and finally the  fifth row displays rates right after the event, indicating a dynamic change of reaction rates during the peak-flow event. 

The spatial and temporal variations of the reaction rates within the streambed vary considerably over the duration of the event. The first row illustrates the rates during the first hour of the event when reactivity rates are low since the system only begins to be dynamic with flow and solute concentrations. The results indicate that for the scenarios with $S = 0.05$ and $\kappa = 10^{-11}$, high reactivity occurs primarily along the deepest subsurface flow paths as well as the stagnation points. Since the results are obtained with respect to the baseflow conditions, we observe that in the regions of stagnation points, when close to the peak-flow time, the DOC consumption and denitrification are lowered. However, this behaviour changes sometime after the peak is attained. On the other hand, for scenarios with $S = 0.001$ and $\kappa = 10^{-12}$, high reactivity occurs closer to the sediment water interface as well as along the deep subsurface flow paths, particularly evident during and after peak-flow conditions. These results demonstrate that under different hydrogeological conditions, removal rates can occur at different depths along the subsurface flow paths for the same peak-flow event. Depending from the event-specific conditions, high reactivity may be encountered at different depths of the streambed as the availability of favourable conditions varies significantly under variable flow and solute input conditions. It is also evident  from Figure 3 and 4 that the flow and transport processes can have differing sensitivities to transience \cite{singh2019dynamic}. This is due to consideration of diffusion-dispersion processes in the reactive model which are enhanced by transient flow conditions \cite{gomez-velez_flow_2017}. 

Studies  (e.g. \cite{harvey2013hyporheic, boano2014hyporheic}) have shown that denitrifical and DOC consumption due to hyporheic exchange exhibit significant variations at meter or even sub-meter scales. The spatial variability depends on the differences in biogeochemical layering of water and solutes such as nitrate, DOC and dissolved oxygen moving along distinct subsurface flow paths. Efficient denitrification requires the following conditions: 1) anaerobic conditions as otherwise oxygen has a higher affinity to act as electron acceptor during respiration and 2) availability of labile DOC. Peak-flow events can drive variable amounts of concentrations of different solutes within different depths of streambed. However, it is important to note that the patterns and dynamics of the contributions of the nutrients fluxes is not necessarily proportional to the events and is dependent on factors such as land-use factors, seasons and sediment-type \cite{blaen2017high} and usually not taken into consideration.

\subsection{Impacts of channel gradient, streambed permeabilities and topography on solute concentration in sediment domain}

\begin{figure}[h]
\centering
\noindent\includegraphics[width=\textwidth]{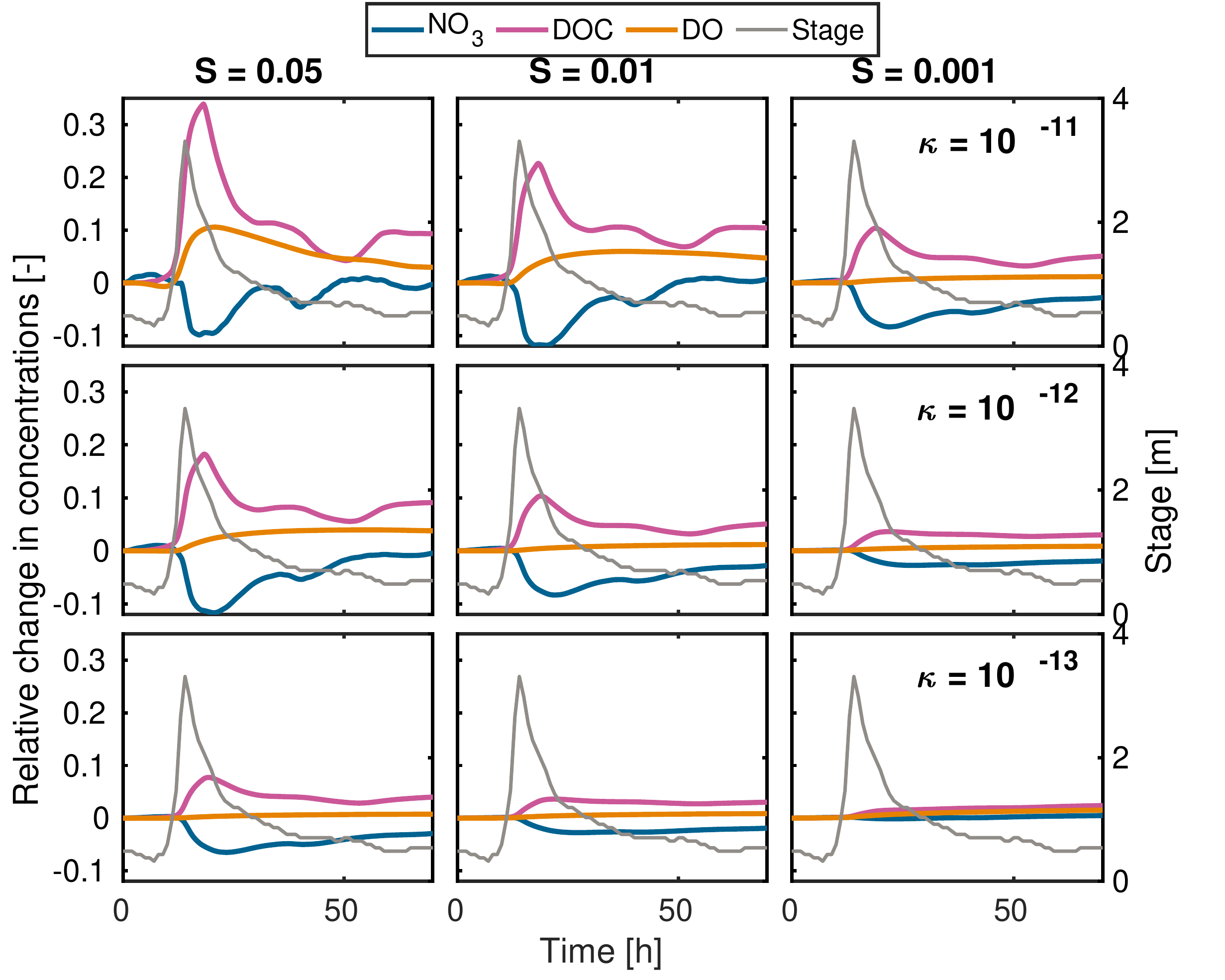}
\caption{Temporal evolution of the relative change in concentrations of nitrate (blue), dissolved organic carbon (pink) and dissolved oxygen (orange)  as compared to baseflow conditions, as a function of time in the sediment domain. The grey solid line indicates the shape of the hydrograph for the event 4 with channel gradient (proportional to channel slope) decreasing (0.05, 0.01 and 0.001) from left to right and  streambed permeabilities  decreasing ($10^{-11}$, $10^{-12}$, $10^{-13}$) from top to bottom.}
\end{figure}

\begin{figure}[h]
\centering
\noindent\includegraphics[width=\textwidth]{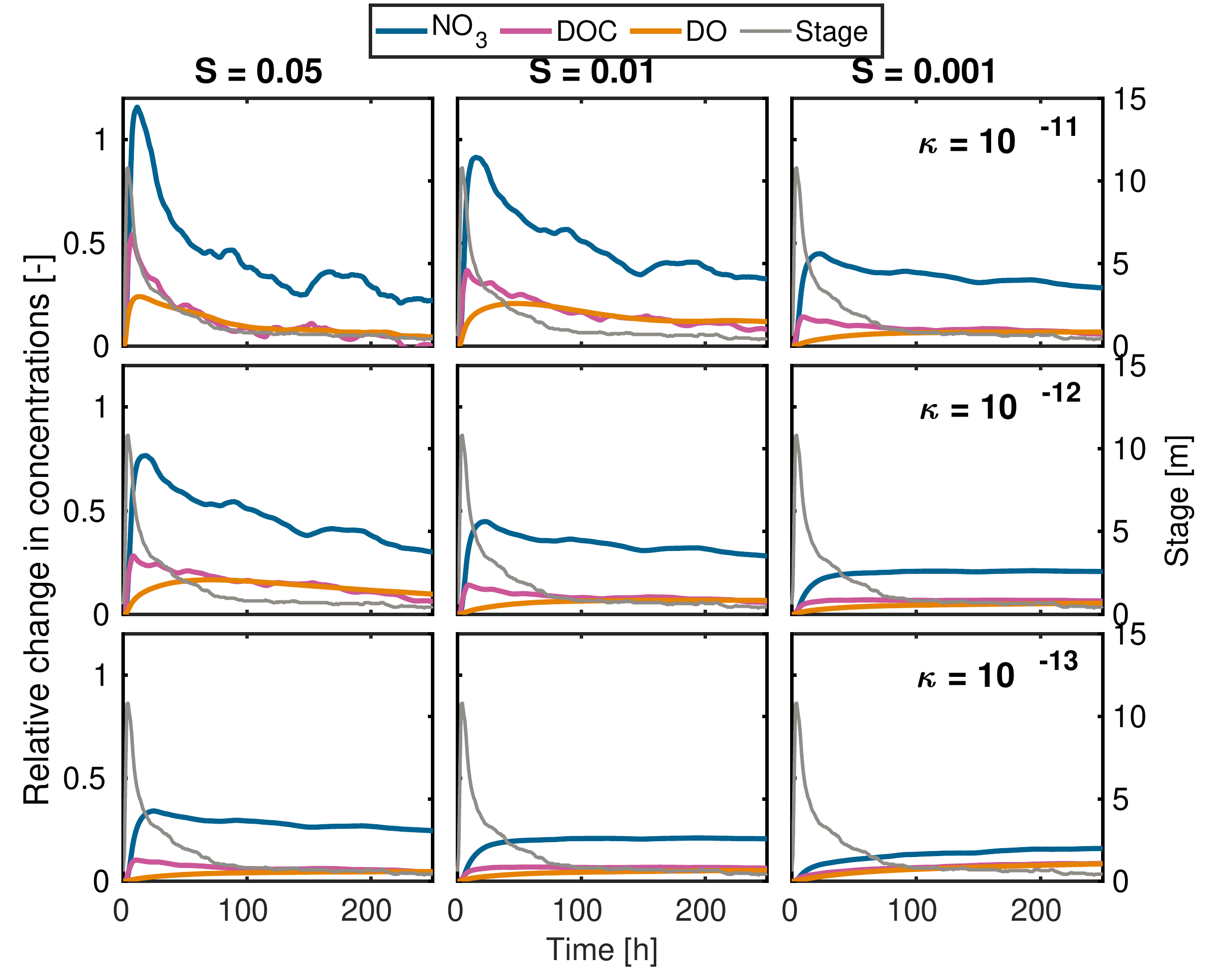}
\caption{Temporal evolution of the relative change in concentrations of nitrate (blue), dissolved organic carbon (pink) and dissolved oxygen (orange)  as compared to baseflow conditions, as a function of time in the sediment domain. The grey solid line indicates the shape of the hydrograph for the event 5 with channel gradient (proportional to channel slope) decreasing (0.05, 0.01 and 0.001) from left to right and  streambed permeabilities  decreasing ($10^{-11}$, $10^{-12}$, $10^{-13}$) from top to bottom.}
\end{figure}

\begin{figure}[h]
\centering
\includegraphics[width=\textwidth]{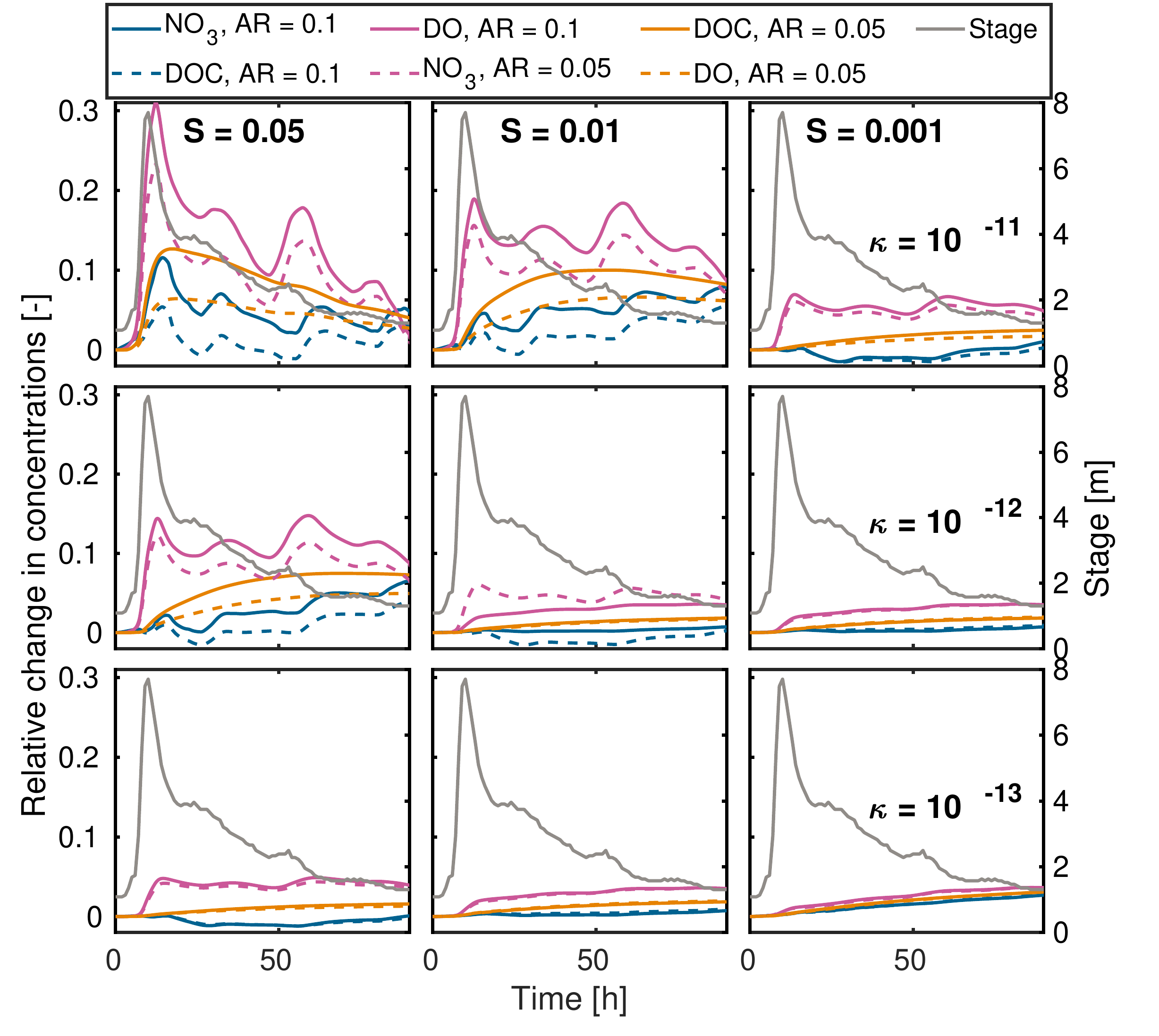}
\caption{Temporal evolution of the relative change in concentrations of nitrate (blue), dissolved organic carbon (pink) and dissolved oxygen (orange)  as compared to baseflow conditions, as a function of time in the sediment domain.  The grey solid line indicates the shape of the hydrograph for the Event 2.  The channel gradient (proportional to channel slope) is decreasing (0.05, 0.01 and 0.001) from left to right and streambed permeabilities are decreasing ($10^{-11}$, $10^{-12}$, $10^{-13}$) from top to bottom. The solid coloured lines represent the scenarios with aspect ratio of 0.1 and dashed lines with aspect ratio of 0.05.}
\end{figure}

Geomorphological properties of the streambed have a considerable impact on nutrient concentrations in the streambed during both the events (Figure 5 and 6). Nitrate concentrations within the streambed domain is reduced to below baseflow conditions for all scenarios, i.e., with varying channel gradient and streambed permeabilities for the Event 4. However, for the Event 5, which is larger in both duration and magnitude compared to Event 4, depicts above baseflow concentrations. In terms of the input boundary concentrations, Event 4 is with an increasing trend in dissolved organic carbon (DOC) concentrations with onset of the storm event however nitrate concentrations with a decreasing  trend. In the case of Event 5 for both DOC and nitrate the concentrations almost follow the trend of the storm event. Please note that stream dissolved oxygen (DO) concentrations for all the events is constant. For Event 1 and Event 3 (not shown) both display a general increasing input of DOC whereas nitrate observed a decreasing trend.   

Scenarios with higher slopes  (i.e. 0.05) and larger sediment permeability ($10^{-11}$) were associated with higher temporal variability in nitrate and DOC concentrations. Both geomorphological settings when combined, due to their large underflow velocities and higher hydraulic conductivity, lead to considerable variations in flow and transport of nutrients in the streambed. Increased underflow velocity causes the hyporheic zone not to expand and causes faster interactions between surface water and ambient groundwater flow, even with high local pressure profiles caused by Event 5. The solute transport is therefore constrained to shallow fast-flowing subsurface circulation paths.

The scenario with a low magnitude and a short duration (Event 4) primarily led to dilution of nitrate in the hyporheic zone because of the decrease of nitrate influx into the streambed with increasing stream discharge. Furthermore, the increased availability of DOC  during this event leads to a rise in both aerobic respiration and denitrification as both reactions require supply of DOC. Contrary to the Event 4, Event 5 is the scenario with a high magnitude and a long duration and is characterized by concentrations of nitrate and DOC following the event, i.e., peaked concentrations coincide with peak-flow of the event. This is because the relative change in streamﬂow during peak-flow event was substantially higher than the associated change in solute concentration. Our results highlight that that the discharge-concentrations relationships can strongly determine the transport and availability of solutes in the streambed.

In addition to the shape of peak-flow hydrographs and associated solute fluxes, the transport of the water and solutes in the streambed is furthermore affected by streambed topography (Figure 7). For diﬀerent bedform aspect ratios of the streambed, the spatial distribution of the hydraulic head at the sediment-water interface varies. Under the influence of transient flow conditions, for high streambed aspect ratios (i.e. bedform amplitude) the effects on nutrient distribution in the sediment domain are enhanced as the local pressure is amplified due to larger advective pumping effect. As shown in Figure 7, the temporal change in nutrient concentrations are generally similar for both aspect ratios (AR), however the change in concentrations is enhanced for higher AR compared to low AR. 

Our results demonstrate that spatial and temporal distribution of  concentrations of nitrate and dissolved organic carbon within the streambed is highly dependent on the changes in local pressure profiles caused by the streambed topography, ﬂow intensity and ambient groundwater ﬂow. Our results clearly indicated greater magnitude of solutes for higher aspect ratio of streambed, higher slope values as well as higher streambed permeability. Higher streambed aspect ratios induce a higher pressure head at the sediment-water interface ($\Delta/H_s$) resulting in deeper streambed subsurface flow paths. On the other hand, steeper streambed slopes induce higher flow velocities across lateral boundaries leading to faster exchange between surface water and ambient groundwater flow. Higher permeabilities of streambed sediment which affect both, the horizontal ambient groundwater flow and the infiltration potential of water and solutes in the streambed, also affect nutrient transport in the streambed. These streambed conditions are particularly important to consider during drier periods when the soil moisture conditions are lower causing greater infiltration. 

\subsection{Temporal variability and hydrogeological controls on hyporheic denitrification potential}

\begin{figure}[h]
\centering
\includegraphics[width=\textwidth]{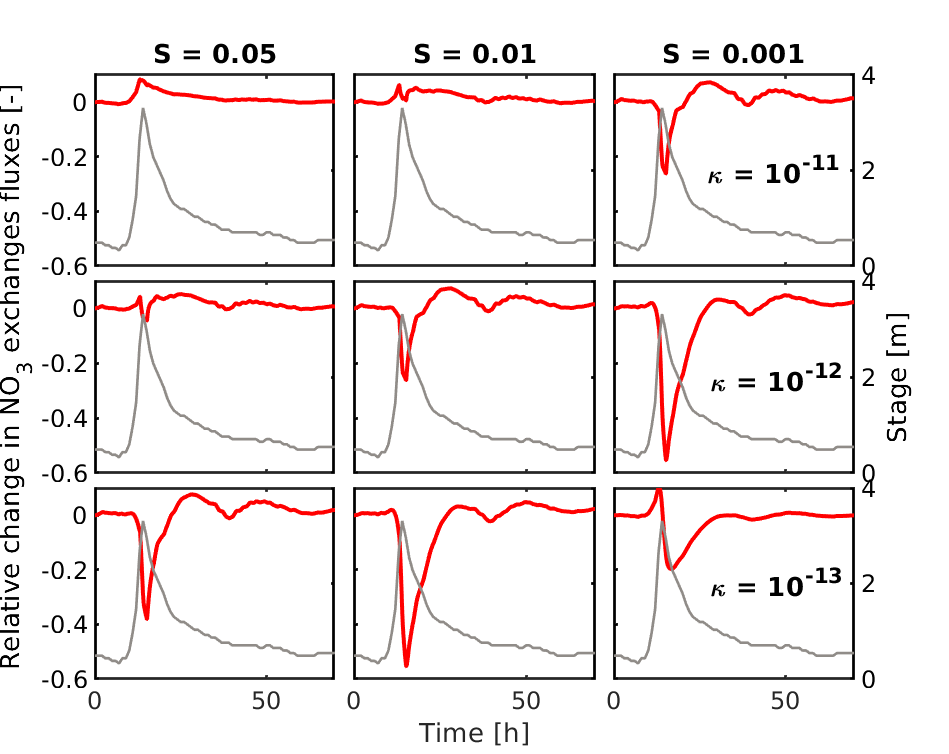}
\caption{Temporal dynamics of relative changes in nitrate concentration fluxes  with respect to baseflow conditions, as a function of time for event 4. The grey solid curve corresponds to the shape of the hydrograph. From the left to the right the channel gradient (proportional to channel slope) is decreasing (0.05, 0.01 and 0.001) and from the top to the bottom streambed permeability values are decreasing ($10^{-11}$, $10^{-12}$, $10^{-13}$). }
\end{figure}

\begin{figure}[h]
\centering
\includegraphics[width=\textwidth]{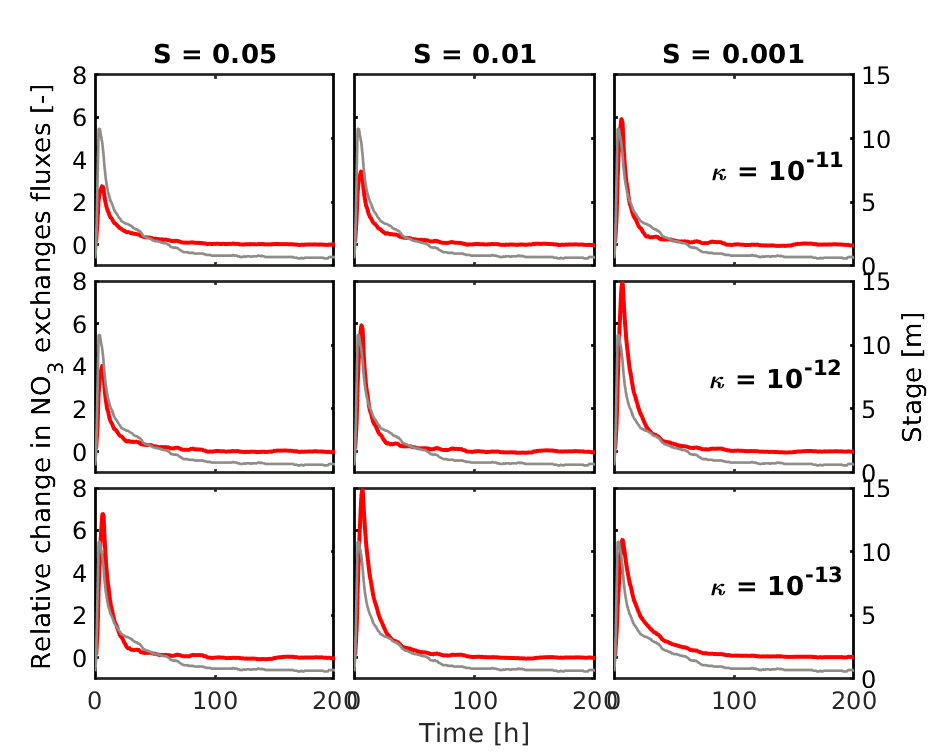}
\caption{Temporal dynamics of relative changes in nitrate concentration fluxes  with respect to baseflow conditions, as a function of time for event 5. The grey solid curve corresponds to the shape of the hydrograph. From the left to the right the channel gradient (proportional to channel slope) is decreasing (0.05, 0.01 and 0.001) and from the top to the bottom streambed permeability values are decreasing ($10^{-11}$, $10^{-12}$, $10^{-13}$). }
\end{figure} 

In order to compare to estimate hyporheic nitrate removal potential for the considered events, we estimated the relative change in nitrate exchange (i.e. difference between influx and outflux) fluxes during the course of the event. The influx and outflux of nitrate is strongly associated with the channel slope, permeabilities as well as the event hydrograph shape along with input solute concentrations (Figure 8 and 9). In Event 4, we observe that the difference between the influx and outflux of nitrate is positive, primarily for certain geomorphological settings ($S = 0.05, 0.01$ and $\kappa = 10^{-11}$). In fact, for conditions with lower streambed permeability, the outflux is higher compared to the influx. However, if we compare these results to Event 5, different behaviour can be observed, i.e., the influx of nitrate is higher than the outflux. This implies that nitrate entering the streambed are largely reduced to molecular nitrogen. The difference between the outflux and influx is largely increasing with decreasing slopes and the relative change is decreasing with decreasing permeabilities. We can also observe that the width between the rising and recession limb for the nitrate exchange fluxes is widening for scenarios with lower permeability values.

Furthermore, we estimated the cumulative nitrate removal and is given by the following mathematical statement
\begin{equation}
R_{NO_3}(t) =  \frac{\int_{0}^t (m_{in} - m_{out}) \ \mathrm{d}t}{M_{in}}
\end{equation}
\noindent where $M_{in}$, the total mass that enters through inflow boundary at the sediment-water interface ($\partial\Omega_{IN}$) and $m_{out}$ is the mass leaving the hyporheic zone through the outflow boundary $\partial\Omega_{OUT}$).

\begin{figure}[h]%
\centering
\includegraphics[width=\textwidth]{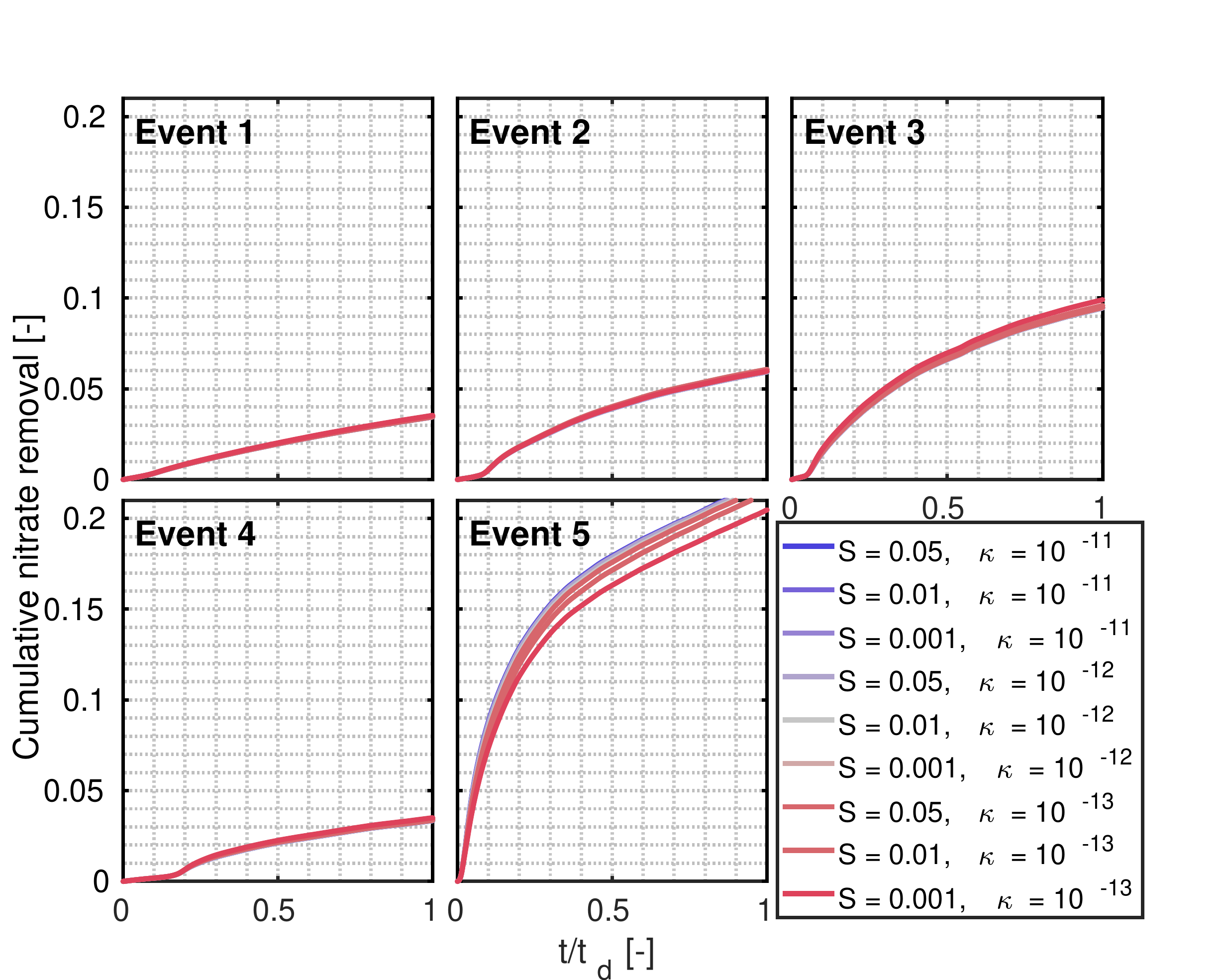}
\caption{Cumulative nitrate concentration with respect to baseflow conditions, as a function of time for the all the considered events and geomorphological settings.  }
\end{figure}

Observations from the breakthrough analysis provided further into the complex reactive transport behaviour on nitrate within a sediment domain. It is evident from the results that shape of the peak-flow event is a dominant factor in the denitrification process (Figure 10). Highest cumulative removal rate is observed for the Event 5 which is the longest duration event. This is followed by the Event 3 with a duration of 25\% less than the Event 5. This is in line with the findings of \citet{singh2019dynamic} where hyporheic zone efficiency, a proxy for the oxygen consumption and denitriﬁcation potential was used, and it was demonstrated that the events with longer duration and higher intensity increase the potential of denitrification. Furthermore, it is noteworthy to observe that the geomorphological settings only become dominant factors for the events with long duration and high intensity and do not play a crucial part in nitrate removal for other events. These ﬁndings are similar to \citet{trauth2017single} where the simulation results for an in-stream gravel bar showed higher rates of aerobic respiration and denitriﬁcation for higher peak-ﬂow intensities and longer duration of the event. Additionally, in particular, scenarios with higher channel slope values and higher permeability depicted higher cumulative nitrate removal. The latter is also in line with the findings of \cite{hester2019effect} where the authors showed that the increasing hydraulic conductivity increased mixing-dependent denitrification.

The denitriﬁcation in anaerobic zones of the HZ is limited by the supply of labile DOC \cite{zarnetske2011dynamics}. During peak-flow events, the availability of dissolved organic carbon in deeper parts of streambed can be increased, which is a requisite for the denitrification process. As favourable conditions may be formed deeper within the streambed during peak-flow events, transported labile DOC supports denitrifying bacteria to complete the process of denitriﬁcation. However, our results indicated that any peak-flow event alone cannot induce increased denitrification, instead specific flood characteristics combined with favourable geomorphic conditions within the streambed determine hyporheic denitrification potential.

\section{Conclusions and outlook}

The denitrification potential under the influence of peak-flow events was studied using a two-dimensional transient reactive transport model. Distinct peak-flow events with characteristic peak-flow intensity, duration and shape were selected from high frequency observations. The impacts of peak-flow events on the denitrification potential were evaluated by considering a range of geomorphological settings, defined by combinations of sediment permeability, channel slope and bedform amplitude.  In addition, we showed that the denitrification potential is sensitive to the input nutrient fluxes and high frequency field observations are important for improving the predictions of denitrification efficiency. Spatio-temporal variations of the flow field and reaction rates within the streambed varied significantly during the course of the peak-flow event. Higher reactivity primarily occurred within deeper subsurface flow paths and near stagnation zones. Furthermore, the the average positive or negative nitrate removal capacity of hyporheic zones was strongly associated with the values of channel slope and permeabilities. Specifically, the hydrograph characteristics emerged as the dominant control for the nitrate removal efficiency, where events with longer duration and higher intensity promoted higher efficiency. The geomorphological parameters additionally modulate the magnitude of the nitrate removal efficiency, such that within the longer duration and higher intensity events, the scenarios with higher slope and permeability values led to marginally higher magnitudes of removal efficiency. 

Peak-flow events are ubiquitous, occurring naturally due to intense precipitation and snow melting, and also due human activities such as periodic release from dam operations and waste water treatment plants. Peak-flow events originating from human activities are typically controlled and hence any cascading effect on water quality or aquatic life can be predicted and managed. However, the natural factors of peak-flow events can be difficult to predict and can have significant implications on overall ecosystem functioning. Our results improve the understanding of the nitrate removal capacity of hyporheic zones during peak-flow events and therefore help predict the after-effects of such events on chemical status of the rivers which can be beneficial for river restoration managers for designing locally optimized restoration measures. Therefore, results presented in this study can be relevant to achieve a good aquatic chemical and ecological health as well as to develop strategies for a sustainable, efficient execution of river management and restoration plans during and after peak-flow events.

This study assumes that conditions in the streambed favour complete denitrification. However, depending on the residence times of water and solutes in the streambed, nitrogen cycling can involve intermediate reactions. For instance, incomplete denitrification can lead to release of $N_{2}0$ which is a greenhouse gas and an ozone-depleting substance \cite{briggs2015physical, comer2020seasonal}, and therefore, may be of particular significance for climate change impacts and related peak-flow events. Exploring the scenarios leading to such conditions under peak-flow events need separate attention and will be taken into consideration in our future work.

\section*{Acknowledgments}
This work is supported by the German Research Foundation under the grant WO671/11-1. G.Chiogna is supported by the German Research Foundation under the grant CH 981/4-1. All data required to reproduce the figures in this paper is available through this link \url{https://git.geomar.de/shubhangi-gupta/hyporheic-denitrification-under-peak-flow-events.git}.
%
%
\bibliographystyle{apalike}
\bibliography{Bibliography}

\end{document}